\documentclass[letterpaper,english,letterpaper,english,prl,notitlepage,twocolumn,superscriptaddress]{revtex4-1}
\usepackage[T1]{fontenc}
\usepackage[latin9]{inputenc}
\usepackage{verbatim}
\usepackage{amsmath}
\usepackage{amssymb}
\usepackage{graphicx}
\usepackage{dsfont}
\usepackage{float}
\usepackage{cancel}

\makeatletter

\pdfpageheight\paperheight
\pdfpagewidth\paperwidth

\usepackage{color}
\usepackage[usenames,dvipsnames,svgnames,table]{xcolor}
\usepackage{comment}
\usepackage{ulem}

\definecolor{violet}{rgb}{0.58, 0.0, 0.83}

\usepackage[caption=false]{subfig}
\usepackage{stackengine}

\newcommand{\gae}{\lower 2pt \hbox{$\,
\buildrel{\scriptstyle >}\over {\scriptstyle \sim}\,$}}
\newcommand{\lae}{\lower 2pt \hbox{$\,
\buildrel{\scriptstyle <}\over {\scriptstyle \sim}\,$}}

\newcommand{\hc}{\text{h.c.}}
\newcommand{\comm}[2]{\left[#1, #2\right]}

 \newcommand{\mkaddcomment}[1]{}
 \newcommand{\mkdel}[1]{}
 \newcommand{\mkdelmath}[1]{}

\usepackage{graphicx}
\usepackage{color}
\begin{document}

\title{Many body localization in the presence of a central qudit}

\author{Nathan Ng}
\affiliation{
Department of Physics and Chemistry, University of California, Berkeley, CA 94720, USA}
\affiliation{
Material Sciences Division, Lawrence Berkeley National Laboratory, Berkeley CA 94720, USA
}

\author{Michael Kolodrubetz}
\affiliation{
Department of Physics, The University of Texas at Dallas, Richardson, Texas 75080, USA
}

\date{\today}

\begin{abstract}
We consider a many-body localized system coupled globally to a central $d$-level system. Under an appropriate scaling of $d$ and $L$, we find evidence that the localized phase survives. We argue for two possible thermalizing phases, depending on whether the qudit becomes fully ergodic. This system provides one of the first examples of many-body localization in the presence of long-range (non-confining) interactions. 
%\begin{description}
%\item[Usage]
%\item[PACS numbers]
%\item[Structure]
%\end{description}
\end{abstract}

\maketitle

A fundamental shift in our understanding of non-equilibrium quantum systems has occurred via the discovery of many-body localization (MBL), where sufficiently strong disorder induces stable localization \cite{Basko2006, Oganesyan2007, Pal2010, Nandkishore2015}. MBL generalizes the notion of Anderson localization to the presence of interactions and is widely believed to be the only generic method for breaking the eigenstate thermalization hypothesis  (ETH \cite{Deutsch1991, Srednicki1994, Tasaki1998}) in isolated quantum systems. Since its inception, MBL has been shown numerically for a variety of models \cite{Oganesyan2007, Pal2010}, mathematically proven to exist under minimal assumptions \cite{Imbrie2016}, and been generalized to situations such as time periodic (Floquet) drive \cite{Lazarides2015, Ponte2015prl}, where MBL is particularly important to avoid heating to a featureless infinite temperature state.

MBL is commonly considered for the case of local interactions, with the exception of \cite{Nandkishore2017}, where long-range confining interactions behave 
short-ranged with regards to the relevant degrees of freedom. Absent confinement, long-range interactions generically entangle spatially separated degrees of freedom, destroying the MBL phase. Perhaps the simplest example of this is the central spin-1/2 model, where it was found that a single globally coupled impurity immediately destroys localization in an infinite spin chain for arbitrarily weak couplings \cite{Ponte2017, Hetterich2018}\footnote{As long as interactions are not taken to scale with system size}. One may suspect that this delocalization is generic for non-confining interactions, as a single spin-1/2 represents in some sense the minimal quantum bath providing thermalization.

In this paper, we show that this intuition is incorrect. Specifically, inspired by quantizing the drive degrees of freedom in Floquet MBL, we show that an appropriate limit of a $d$-level system (``qudit'') coupled to a disordered spin chain may display an MBL-ETH transition at finite coupling. We argue that this phase transition survives the thermodynamic limit under the condition that $d \gae \sqrt L$ asymptotically, where $L$ is the length of the spin chain. The resulting phase diagram has many surprising features, such as decreased thermalization for larger $d$ and the potential for an inverted mobility edge.

	\begin{figure}[b]
	\includegraphics[width=\columnwidth]{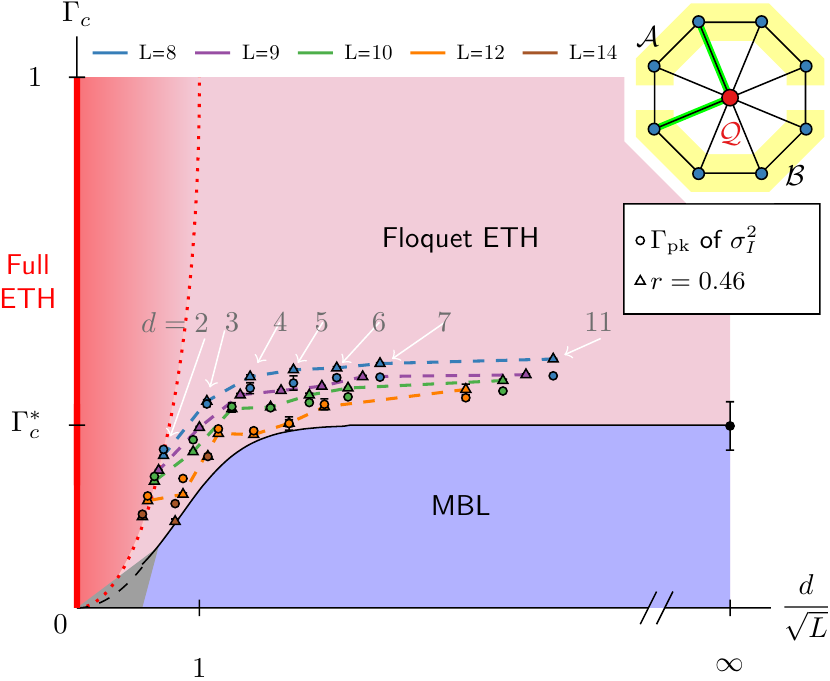}
	\caption{Proposed infinite temperature phase diagram for the central qudit model upon taking $L\to\infty$ with $d/\sqrt{L}$ fixed. In addition to MBL and ETH phases of the spin chain, the dotted line indicates the crossover from fully thermal qudit to athermal qudit (``Floquet ETH''). The behavior of the phase boundary near $d/\sqrt{L} = 0$ is unclear; a possible $\Gamma_c \sim L^{-1}$ scaling \cite{Ponte2017} is indicated by the dashed line. Two finite size estimators of the critical $\Gamma$ -- defined near  Eq.~\ref{eq:mutual_info} and in Fig.~\ref{fig:MIslowDrift} -- are plotted. The value $r = 0.46$ is taken to be halfway between the thermal ($r=0.53$) and non-thermal ($r=0.39$) values \cite{Oganesyan2007}.}
	\label{fig:phaseDiagram}
	\end{figure}

	\begin{figure*}
\includegraphics[width=.9\textwidth]{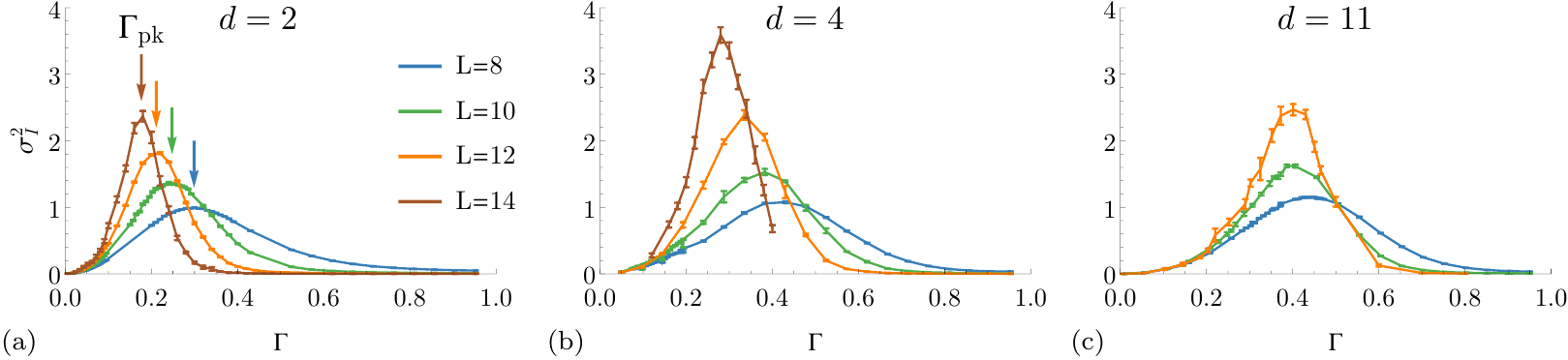}
		\caption{Half chain mutual information variance $\sigma^2_{I}$ for various qudit sizes. The leftward drift of the peaks with system size $L$ is seen to slow down with increasing $d$.}
		\label{fig:MIslowDrift}
	\end{figure*}

\paragraph{Model\textemdash} As a starting point, we consider a model of MBL in the presence of global periodic drive, adapted from Zhang et al.\ \cite{Zhang2016}:
	\begin{align*}
		H &= \frac{ H_z + H_x}{2} + \cos{(\Omega t)} \frac{H_z - H_x}{2} \\
		H_z &= \sum_i (h + g \sqrt{1-\Gamma^2} G_i) \tau^z_i + \tau^z_i \tau^z_{i+1} \\
		H_x &= g \Gamma \sum_i \tau^x_i
	\end{align*} where $\tau$ are Pauli matrices and $G_i$ are random Gaussian variables of zero mean and unit variance describing on-site disorder. When the drive frequency is high, the system is effectively described by the average Hamiltonian $\frac{1}{2}(H_z + H_x)$, which exhibits an MBL-ETH transition. The coupling $\Gamma$ controls the strength of disorder as well as the degree of noncommutativity between the zeroth and first harmonics of $H$. We take $h=0.809$, $g=0.9045$, and $\Omega = 3.927$, for which we numerically verify that an MBL-ETH transition is present at $\Gamma_c \approx 0.33$.

Interesting insight may be obtained by examining this model in the Floquet extended zone picture \cite{Shirley1965}. Writing the wave function in Fourier harmonics, $|\psi(t)\rangle =\sum_{n=-\infty}^{\infty} |\psi^{(n)}(t)\rangle e^{in\Omega t}$, $|\psi^{(n)}\rangle$ may be considered as the wave function dressed by $n$ photons. This wave function evolves under the extended zone Hamiltonian:
	\begin{align}
		\label{eq:extZoneHam}
		H_{\text{EZ}} = & \sum_n \left(  \frac{1}{2}H_+  + \Omega n \right) \otimes |n\rangle \langle n| \nonumber \\ & + \frac{1}{4} H_- \otimes \left( \sum_n |n + 1\rangle \langle n| + \hc \right) ,
	\end{align}
where $H_{\pm} = H_z \pm H_x$ are the zeroth ($+$) and first ($-$) Fourier modes of $H$. We introduce an extended Hilbert space $|n\rangle$ corresponding to photon occupation numbers.

In numerically solving such an extended zone Hamiltonian, one often truncates the photon Hilbert space, for instance restricting $n = -N_c, -N_c + 1, \ldots, N_c$. In order to obtain the proper Floquet result, one must extrapolate $N_c \to \infty$. If instead we maintain a finite truncation, the photons form a $d$-level system -- a ``qudit'' -- with $d=2 N_c + 1$. In the $d\to\infty$ limit, we recover Floquet physics, for which an MBL-ETH transition is expected in this model. Keeping $d$ finite, as in the case of a qubit ($d=2$), Ponte et al.\ have argued in a similar model that ETH is expected for all finite couplings in the thermodynamic limit \cite{Ponte2017}. The remainder of this paper will be devoted to understanding the crossover between these limits, thereby uncovering the physics of MBL in the presence of a central qudit. Note that alternative choices of truncation would allow one to instead think of a central spin-$S$ or photon with finite occupation in place of the qudit, a picture relevant to cavity QED. These other truncations are  discussed in the Supplement Material \footnote{See Supplemental Material URL for analysis of this system within qudit and photon truncations, as well as additional numerical details. The supplement also includes Refs.\ \cite{Goldman2014, Bukov2016,Weinberg2017, Keating2015, Wilming2017}}.

\paragraph{Numerical results\textemdash}
	We investigate the behavior of this model up to $L=14$ spins and $d = 11$ using the shift-invert method \cite{GolubVanLoan}. By targeting the ten states with energy closest to 0, we effectively work in the infinite temperature limit. We see that these ten states describe the same energy density by observing that there are no small scale structures in the disorder-averaged many-body density of states near zero energy \cite{Kim2014, Garrison2018}. We compare these results to the full Floquet dynamics ($d=\infty$) by approximating the exact dynamics over one period with $\geq 16$ time steps.

In our model, thermalization of the localized spins can occur through direct spin-spin interactions, qudit-mediated interactions, or some combination thereof. To distinguish entanglement between the spins from entanglement with the qudit, we consider the mutual information (MI) between two halves of the spin chain (see Fig. \ref{fig:phaseDiagram} for definition of $\mathcal A$ and $\mathcal B$):
\begin{align}
  I(L/2) \equiv I(\mathcal{A},\mathcal{B}) = S(\rho_\mathcal{A}) + S(\rho_\mathcal{B}) - S(\rho_\mathcal{AB}). \label{eq:mutual_info}
\end{align}
By subtracting entanglement with the qudit, $S(\rho_\mathcal{AB}) = S_{\text{qudit}}$, we find that $I$ captures the bipartite correlations between $\mathcal{A}$ and $\mathcal{B}$ more faithfully than $S(\rho_{\mathcal{A}})$.

%\cite{Page1993}

	\begin{figure*}
\includegraphics[width=0.98\textwidth]{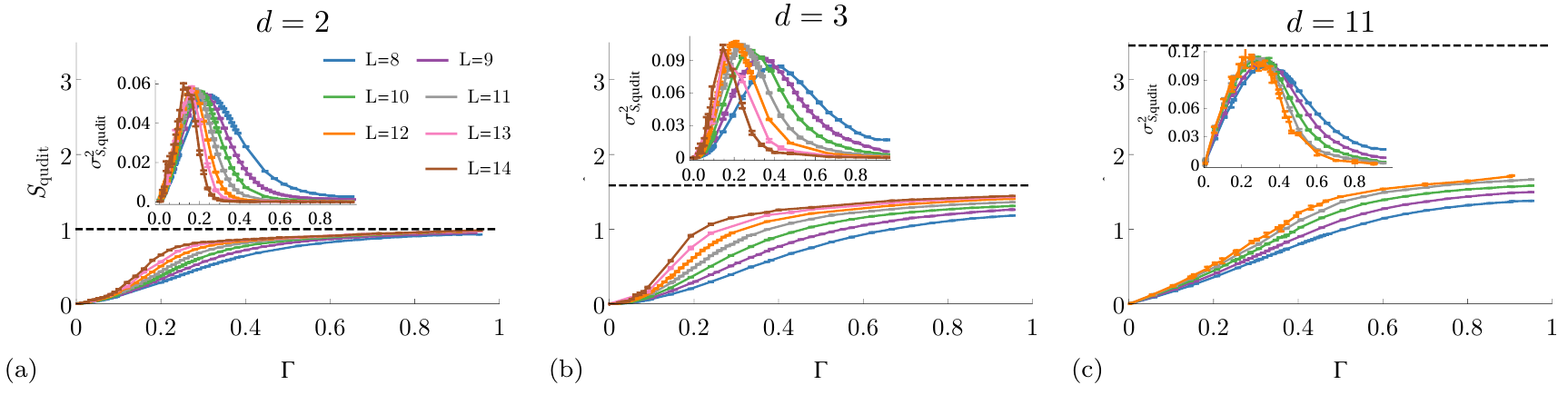}
\caption{Entanglement entropy $S_{\text{qudit}} \equiv S(\rho_{\mathcal{Q}}) = S(\rho_{\mathcal{AB}})$ between the qudit and the spin chain. The dashed line corresponds to the Page value, $\log_2{d} - d\left({2^{L+1}\log{2}} \right)^{-1} \approx \log_2{d}$. The insets show the variance of $S_{\text{qudit}}$.}
\label{fig:qdsavg}
\end{figure*}

        We calculate mutual information and qudit entanglement entropy for each of the eigenstates and $200 - 6000$ realizations of disorder, as well as the level statistics ratio $r$ \cite{Oganesyan2007}. Let us begin by discussing $I(L/2)$. For large $d=11$ approaching the Floquet limit, it increases from a nearly system size independent area law in the MBL phase at small $\Gamma$ to a thermal volume law, approaching the Page value $\displaystyle S_{\text{Page}} = \frac{L}{2} - \frac{1}{2 \log{2}}$ \cite{Page1993}, for large $\Gamma$ (see Supplement \cite{Note2}). It has been found elsewhere that shot-to-shot fluctuations of the entanglement entropy are a useful detector of the MBL-ETH phase transition, peaking sharply near the transition \cite{Kjall2014, Khemani2017}. Here we obtain the variance of the MI, $\sigma_I^2$, due to intersample variations between disorder realizations and intrasample variations between eigenstates (Fig. \ref{fig:MIslowDrift}). Treating the peak values $\Gamma_{pk}(L)$
%        \mkadd{--- denoted as $\Gamma_{pk}(\sigma^2_I)$ in Fig. \ref{fig:phaseDiagram} ---}
        as a finite size approximation of the critical point, we see that for large $d$, the peak shifts only weakly with $L$. This is consistent with the Floquet MBL-ETH phase transition at finite $\Gamma$ in taking first $d\to \infty$, then $L \to \infty$. By contrast, at the smallest value of $d=2$, the peak shifts sharply with $L$, consistent with the expected absence of an MBL-ETH phase transition in the thermodynamic limit. The behavior for $d\sim 5$ is intermediate to these two limits, and its crossover behavior will be addressed in more detail later.

The qudit entanglement entropy $S_{\text{qudit}}$ and its variance, $\sigma^2_{\text{S,qudit}}$, are shown in Fig. \ref{fig:qdsavg}, while $r$ is shown in the Supplement \cite{Note2}. One striking difference between $S_\mathrm{qudit}$ and $I(L/2)$ is immediately apparent -- for large $d$, the qudit entropy does not reach its maximal value, and thus the qudit does not thermalize. Despite the lack of thermalization in the qudit, the level statistics ratio still saturates the Gaussian orthogonal ensemble value of $r\approx 0.53$ for $\Gamma > \Gamma_c$ in the large $d$ limit. On the other hand, for $d=2$, the qudit entropy and its fluctuations closely track $I(L/2)$, suggesting that thermalization of the spin chain is mediated by the central qudit. These numerics together suggest that thermalization of the qudit and the spin chain do not always go hand in hand, confirming the expectation that the limits $d\to\infty$ and $L\to\infty$ do not commute. We now address how these limits may be taken to obtain the phase diagram shown in Fig. \ref{fig:phaseDiagram}.

The appropriate scaling of $d$ vs. $L$ can be argued by first decoupling them, i.e., taking $\Gamma=0$. Then eigenstates of the full problem become direct products of eigenstates of $H_z$ with those of the qudit. The qudit states behave like non-interacting charged particles in an external electric field with nearest neighbor hopping proportional to the many-body energy of the $H_z$ eigenstate. In the Floquet limit, $d\to\infty$, the qudit will be Wannier-Stark localized with a characteristic spread given by the ratio of the hopping strength $\langle H_z \rangle$ to the potential tilt $\Omega$, for which the variance of the qudit occupation is given by $\Delta_Q^2 \equiv \langle n^2 \rangle - \langle n \rangle^2 = \frac{1}{2} \langle H_z \rangle^2 / \Omega^2$ \cite{Note2}. The many body spectrum has characteristic width $\sigma_{\langle H_z \rangle} \sim \sqrt L$, hence averaging over eigenstates gives $\Delta_Q^2 \sim L$.

\begin{figure}%[h]
\includegraphics[width=0.8\columnwidth]{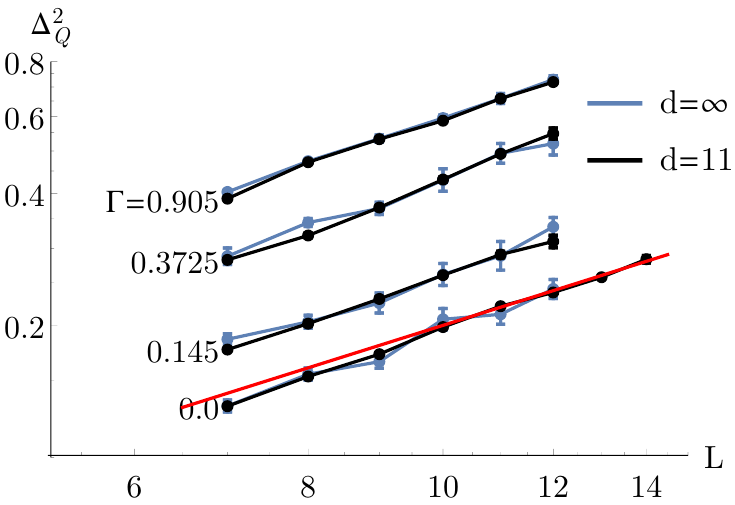}
\caption{Variance of the qudit wavefunction versus size of the spin chain, for $\Gamma$ in both MBL and ETH phases and near criticality. The red line is given by $0.02 L$.
}
\label{fig:qdVarLogLog}
\end{figure}

This scaling of $\Delta_Q^2$ is further argued to be robust for small $\Gamma$ in the Supplement \cite{Note2}. However, numerically we find that this result holds nonperturbatively as well, giving $\Delta_Q^2 \sim L$ for $\Gamma$ throughout the phase diagram (Fig. \ref{fig:qdVarLogLog}). Therefore, we argue that the relevant ratio controlling thermalization is $d/\sqrt L$, as in Fig. \ref{fig:phaseDiagram}. For $d \gg \sqrt L$, the spin chain is insufficient to act as a bath for the qudit, and thus no thermalization of the qudit occurs. For $d \ll \sqrt L$, the spin chain can thermalize the qudit and vice versa. Taking the limit $L \to \infty$ with $d/\sqrt L$ small but finite, our data is unable to confirm whether the qudit fully thermalizes, or rather whether the qudit entropy gradually crosses from athermal to thermal as we take $d/\sqrt L \to 0$; we leave this topic for future study.

Having identified $d/\sqrt L$ as the relevant scale for understanding the qudit's role in thermalization, we may now plot the finite size approximants to $\Gamma_c$ (Fig. \ref{fig:phaseDiagram}). We see that once $L$ is reasonably ``large'' ($L \gae 10$) the finite size $\Gamma_c$ from level statistics and MI variance seem to approach a single curve, which we postulate will become a sharp MBL-ETH phase transition in the thermodynamic limit. For $d/\sqrt L \lae 1$, the MBL-ETH transition indicated by these two measures is consistent with that obtained from the qudit entanglement entropy, while going to $d/\sqrt L \gae 1$, this is no longer true, consistent with a crossover from qudit-mediated thermalization \cite{Note2}. Finally, we note that the prediction of $\Gamma_c \sim 1/L$ at arbitrary finite $d$ \cite{Ponte2017, Hetterich2018} maps in our phase diagram to $\Gamma_c \sim d^2 / L$ for $d/\sqrt L \ll 1$. We are unable to obtain data for transitions in this limit, so leave clarification of the bottom left corner of the phase diagram for future work.
	
\paragraph{Discussion\textemdash}

Our data suggest that three distinct phases exist for the disordered spin chain coupled to a central qudit: (1) Both spin chain and qudit are athermal (MBL), (2) both the spin chain and the qudit are thermal (full ETH), and (3) the spin chain is thermal but the qudit is athermal. We refer to this last phase as Floquet ETH because it is necessarily obtained in the Floquet limit, $d/\sqrt L \to \infty$. By contrast, for full ETH to occur, the spin chain must act as a bath for the qudit states and vice versa. In the thermodynamic limit, this should manifest as observables for both the spins and the qudit exhibiting criticality at the same value of $\Gamma$. We cannot currently probe this effect, given the small region of $L$ and $d$ space accessible. However, drifts in $\Gamma_\mathrm{pk}$ obtained from $S_\mathrm{qudit}$ (Fig. \ref{fig:qdsavg}, inset) and $I(L/2)$ (Fig. \ref{fig:MIslowDrift}) appear to be consistent with a $\Gamma_c = 0$ transition as found in earlier works \cite{Ponte2017, Hetterich2018}. The full ETH phase is certainly obtained for $d/\sqrt L=0$, e.g., by taking $L\to\infty$ while keeping $d$ finite. While we cannot rule out the possibility that this phase extends to nonzero $d/\sqrt L$, implying a phase transition between the thermal full ETH and Floquet ETH phases, we expect that Floquet ETH will be immediately obtained as $d/\sqrt L$ is increased from zero.

Most surprising is the persistence of MBL at finite $d/\sqrt L$. Integrating out the central qudit, we may think of this as MBL in the presence of infinite range interactions. Similar MBL phases have been proposed in the presence of long-range confining interactions by Nandkishore and Sondhi \cite{Nandkishore2017}, but this work represents the first numerical example of such long-range-interacting MBL to our knowledge. A natural expectation is that thermalization would be easier for larger central qudit size, as larger central qudits have more pathways for the qudit to flip and thus mediate long-range interactions. However, our data suggests the opposite -- larger $d$ leads more readily to MBL. In the Supplement, we show how the qudit may be ``integrated out'' in the high frequency limit and recover the $d/\sqrt L$ scaling using this method. Intuitively, the picture that emerges is that, at large $d$, the spin chain Hamiltonian becomes independent of the qudit occupation due to translation invariance in qudit occupation space, and thus the qudit \footnote{Or photon, or central spin} is no longer able to mediate long-range interactions. Finally, we note that applying the same procedure to models with Floquet-induced localization \cite{Bairey2017, Choi2018} would lead to localization that is encouraged rather than discouraged by the presence of the central qudit.

We note one further non-trivial corollary to this phase diagram. If we treat $d$ as a proxy for the photon number in a photonic regularization of the Floquet problem, then smaller $d$ would correspond to smaller photon number and, thus, lower many-body energies. Moving to the left in Fig. \ref{fig:phaseDiagram} is then loosely equivalent to decreasing energy. If we take some value of $\Gamma$ below the Floquet critical point, e.g., $\Gamma=0.2$, this implies that the system goes from many-body localized at infinite temperature to ergodic at lower temperature: an inverted many-body mobility edge. This analogy is inexact, but numerically we may target lower energy densities at fixed $d$ to determine whether indeed this unexpected inversion holds.

Experimentally, central qudit systems are realized in a variety of settings, such as quantum dots \cite{Kikkawa1998, Khaetskii2002} and defect centers \cite{Hanson2008}. Localization of the spin bath there is less obvious, as the spin-spin interactions are commonly dipolar. Other promising avenues for realizing localization in the presence of a central mode include superconducting qubits coupled in geometry similar to Fig.~\ref{fig:phaseDiagram} \cite{Roushan2017} or spin chains consisting of ultracold atoms globally coupled to a cavity \cite{Brennecke2007, Kubo2010, Amsuss2011, Klinder2015}. In the latter architecture, the cavity photon number plays the role of the qudit size, as discussed more extensively in the Supplement \cite{Note2}.

In conclusion, we have mapped out the phase diagram of a disordered spin chain interacting with a central qudit. We found that the size of the central qudit plays an important role, with the ratio $d / \sqrt L$ appearing to control the crossover from Floquet-like physics at $d / \sqrt L \gg 1$ to central qudit-like physics at $d / \sqrt L \ll 1$. We expect similar behavior to hold for other models of Floquet MBL, as well as other methods for quantizing the Floquet drive.

\paragraph{Acknowledgments\textemdash}

We would like to acknowledge valuable discussions with Marin Bukov, Anushya Chandran, Greg Meyer, Rahul Nandkishore, Zohar Nussinov, Anatoli Polkovnikov, Maksym Serbyn, Bj\"{o}rn Trauzettel, and Romain Vasseur. We also acknowledge support from the U.S. Department of Energy Basic Energy Sciences (BES) TIMES initiative and UTD Research Enhancement Funds. This research used resources of the National Energy Research Scientific Computing Center, a U.S. Department of Energy Office of Science User Facility operated under Contract No. DE-AC02-05CH11231.

\bibliography{references}

\newpage
\onecolumngrid

\appendix
\section{Perturbative argument for $d \sim \sqrt{L}$ scaling at small $\Gamma$}

In the main text, we have argued that the variance of the qudit wave function scales as $\Delta^2_{\mathcal Q} \sim L$ for $\Gamma=0$, leading to $d \sim \sqrt{L}$ scaling. Here we provide more details and extend this argument perturbatively to small $\Gamma$.

To lowest order in $\Gamma$, the Hamiltonian (Eq. 1 of main text) 
%\eqref{eq:extZoneHam} 
takes on the form:
\begin{align*}
  H &= \sum_{n=-\infty}^{\infty} (H'_z/2 + \Omega n) |n\rangle \langle n| + \frac{H'_z}{4} (|n \rangle \langle n+1| + \hc) + \Gamma \left( \sum_{i=1}^{L} \tau^x_i \right) \left( \frac{|n\rangle \langle n|}{2} + \frac{|n \rangle \langle n+1| + \hc}{4} \right), 
\end{align*}
where $H'_z = \sum_i \tau^z_i (h + g G_i) + \sum_i \tau^z_i \tau^z_{i+1}$. For $\Gamma = 0$, the solution is given in terms of Wannier-Stark states localized around $|-n\rangle$ (up to normalization):
\begin{align*}
  |\tau, \psi_n \rangle &= |\tau\rangle \otimes \sum_r J_{r+n}\left( \frac{H'_z(\tau)}{2\Omega} \right) |r\rangle ,
\end{align*} where $|\tau\rangle$ is the state of the spin chain, with spectrum $E_n = H'_z(\tau)/2 - n \Omega$, and $J_n$ is the Bessel function of the first kind. $H'_z$ is composed of the sum of $L$ local operators, yielding a distribution of energies $H'_z(\tau)$ with standard deviation $\sim \sqrt{L}$. Meanwhile, the variance of the qudit occupation can be computed exactly:
\begin{align*}
  \Delta^2_{\mathcal{Q}} &= \left[ \sum_{r=-\infty}^{\infty} r^2 J^2_{r}\left( \frac{H'_z(\tau)}{2\Omega} \right)\right] - \left[ \sum_{r=-\infty}^{\infty} r J^2_{r}\left( \frac{H'_z(\tau)}{2\Omega} \right)\right]^2 \\
  &= 2 \sum_{r=1}^{\infty} r^2 J^2_{r}\left( \frac{H'_z(\tau)}{2\Omega} \right) \\
  &= \frac{1}{2} \left( \frac{H'_z(\tau)}{2\Omega} \right)^2
\end{align*}

In finite qudits this expression is no longer valid once $d \sim \sqrt{L}$, as the sums over qudit states are no longer infinite. Assuming that the $\Gamma=0$ eigenstates are continuously connected to the MBL eigenstates at $\Gamma > 0$, we can examine linear order corrections to the eigenstates:
\begin{gather*}
  |\tau, \psi_n \rangle^{(1)} = \Gamma \sum_{m, \tau'}  \left( \sum_{i=-\infty}^\infty \langle \tau' | \tau_i^x | \tau \rangle \right) \frac{f(n-m)}{\Delta E} |\tau',\psi_m\rangle \\
  f(n-m) = \sum_r \frac{J_{r+n}\left( \frac{1}{2\Omega} H'_z(\tau)\right)}{2} \left[ J_{r+m}\left( \frac{1}{2\Omega} H'_z(\tau') \right) + \frac{J_{r+1+m}\left( \frac{1}{2\Omega} H'_z(\tau') \right)+J_{r-1+m}\left( \frac{1}{2\Omega} H'_z(\tau') \right)}{2}\right] ,
\end{gather*}

Because $\tau'$ must be related to $\tau$ by one spin flip, $H'_z(\tau') - H'_z(\tau) \sim O(1)$. Hence when $L$ is large, the arguments of the Bessel functions appearing in $f(n-m)$ are essentially the same, and $f$ essentially becomes the autocorrelation of $J_r$. 
A rough approximation for the Bessel function $J_r(x)$ has 
\[ J_r(x) \approx 
\begin{cases}
  \sqrt{\frac{2}{\pi x}} & |r| < x \\
  0 & \text{otherwise}
\end{cases},
\]
%	non-zero over $|r| < x$ 
for integer $r$ and large arguments.	
With this approximation, $f(n-m)$ is zero when $|n-m| \gtrsim 2 \frac{1}{2\Omega} H'_z(\tau) \sim O(\sqrt{L})$ and is maximal when $n\approx m$. The first order perturbation of $|\tau, \psi_n\rangle$ will only involve $O(2\sqrt{L})$ more qudit spin states, preserving the $d \sim \sqrt{L}$ scaling found seen for $\Gamma = 0$.

Though suggestive, to treat this carefully requires a nonperturbative calculation, which would make clear the existence of a finite $\Gamma$ MBL transition. 

\section{$d\sim \sqrt L$ scaling from high frequency expansion}

In this section, we provide an alternative analytical approach to the $d \sim \sqrt L$ scaling by ``integrating out'' the central mode. We begin by reintroducing time dependence, enabling use of a high frequency expansion. We show how this generically provides a simple effective Hamiltonian in which the central mode enters only through its occupation number, provided that the high frequency expansion is convergent. The high frequency expansion applies generally for arbitrary choice of central mode, but in subsequent sections we detail its application to two important choices: a central qudit, as addressed in the main text, and a ``central'' bosonic degree of freedom as would be relevant in cavity QED. For both cases, we argue that $d \sim \sqrt L$ scaling naturally emerges. The central mode can additionally be thought of as a central spin-$S$, with $S = (d-1)/2$. Our analysis applies equally well to this case, but we will not consider it in detail.

\subsection{``Integrating out'' the central mode via high frequency expansion}

We first reintroduce time-dependence by going into a rotating frame, \[ |\psi_{\text{rot}}(t) \rangle = e^{i\Omega \hat{n} t} |\psi(t)\rangle. \] Inserting this into the Schrodinger equation, one sees that the tilt $\Omega\hat{n}$ is cancelled, and the states in the rotating frame evolve under the time-dependent Hamiltonian 
\begin{align*}
  H_{\text{rot}}(t) &= e^{i\Omega \hat{n}t} H e^{-i\Omega \hat{n} t} - \Omega \hat{n} \\
  &= \underbrace{\left( \frac{H_z + H_x}{2} \right)}_{H_0} + \underbrace{\left( \frac{H_z - H_x}{4} \right)}_{H_1} \left( e^{i\Omega t} \sigma^+ + e^{-i \Omega t}\sigma^- \right)~,
\end{align*}
where the operators $\sigma^\pm$ and $\hat{n}$ raise/lower the central mode's state and measure the occupation in the central mode, respectively. The detailed form of these operators depending on the truncation scheme and are defined in Eqs.~\ref{eq:operator_qudit} and \ref{eq:operator_photon}. 

In restoring time-dependence we now have a Floquet problem with period $T$, albeit a completely different Floquet problem from that obtained in the conventional $d\to \infty$ Floquet limit. By Floquet's theorem, the time evolution in this rotating frame factorizes into the form
\begin{equation*}
  U_{\text{rot}}(t)=e^{-iK_{\text{rot}}(t)}e^{-iH_{\text{rot}}^{\text{eff}}t}e^{iK_{\text{rot}}(0)},
\end{equation*}
where $H^{\text{eff}}_{\text{rot}}$ is the time-independent effective Hamiltonian which describes long time evolution and $K_{\text{rot}}(t)=K_{\text{rot}}(t+T)$ is the time-periodic kick operator which describes micromotion along with the ``kick'' into the effective frame. In the limit of high-frequency drive ($\Omega \gg J$), the van Vleck high frequency expansion (HFE) gives expressions for $H_\mathrm{rot}^\mathrm{eff}$ and $K_\mathrm{rot}$ order by order in $\Omega^{-1}$ \cite{Goldman2014}. For a drive like ours involving only first harmonics, i.e., $H(t) = H_0 + \overbrace{\left( V^{(1)}e^{i\Omega t} + V^{(-1)}e^{-i\Omega t}\right)}^{V(t)}$, the high frequency expansion gives
\begin{align*}
  H^{\text{eff}}_{\text{rot}} =& H_0 + \frac{1}{\Omega}\comm{V^{(1)}}{V^{(-1)}} + \frac{1}{2 \Omega^2} \left( \comm{\comm{V^{(1)}}{H_0}}{V^{(-1)}} + \hc \right) \\
  &+ \frac{1}{\Omega^3} \left( \comm{\comm{\comm{V^{(1)}}{H_0}}{H_0}}{V^{(-1)}} + \frac{2}{3} \comm{\comm{V^{(1)}}{\comm{V^{(1)}}{V^{(-1)}}}}{V^{(-1)}}+ \hc \right) + O(\Omega^{-4}) \\
  iK_{\text{rot}}(t) =& \frac{1}{\Omega}\left( e^{i\Omega t} V^{(1)} - \hc \right) + \frac{1}{\Omega^2} \left( e^{i\Omega t} \comm{V^{(1)}}{H_0} - \hc \right) \\
  &+ \frac{1}{\Omega^3} \left( e^{i\Omega t} \comm{\comm{V^{(1)}}{H_0}}{H_0} + \frac{e^{2 i \Omega t}}{4} \comm{\comm{V^{(1)}}{H_0}}{V^{(1)}} + \frac{2 e^{i\Omega t}}{3} \comm{V^{(1)}}{\comm{V^{(1)}}{V^{(-1)}}} - \hc \right) + O(\Omega^{-4}).
\end{align*}

When this expansion in $\Omega^{-1}$ converges, one sees that $\comm{\hat{n}}{H^{\text{eff}}_{\text{rot}}} = 0$, since $H^{\text{eff}}_{\text{rot}}$ is time independent. From the form of the first harmonic, $V^{(\pm 1)} = H_1 \sigma^{\pm}$, it is apparent that all-to-all couplings in the spin chain come as powers of $(H_1)^2$ at high orders of the expansion. Similarly, high order terms in the kick operator serve to delocalize the central mode, incrementally coupling it to adjacent states with increasing powers of $\Omega^{-1}$. 

The commutation of $\hat{n}$ and $H^{\text{eff}}$ leads to the following relationships between eigenstates and eigenenergies in the effective rotating frame and the lab frame:
\begin{align}
  |E\rangle_{\text{lab}} &= e^{-i K_{\text{rot}}(0)}|\epsilon_{i}; n\rangle_{\text{rot}}^{\text{eff}} \label{eq:stateRelation} \\
  E &= \epsilon_i(n) + \Omega n \label{eq:framesEnergy},
\end{align} where $\epsilon_i(n)$ labels the eigenenergy of $H_\mathrm{eff}^\mathrm{rot}$ for given $n$. Intuitively, we may see that Eq.~\ref{eq:framesEnergy} follows from continuity. In the extreme high frequency limit $\Omega \gg JL$, where the expansion is absolutely convergent (as opposed to asymptotically convergent), the spectrum of effective energies in the rotating frame, $\epsilon_i(n)$, ranges from roughly $-JL$ to $+JL$. In the lab frame, each branch of this effective Hamiltonian is indexed by its mode occupation number $n$ with energy centered around $\Omega n$. Thus Eq.~\ref{eq:framesEnergy} clearly holds in this extreme limit. As one adiabatically decreases $\Omega$ while maintaining the (at least asymptotic) convergence of the high frequency expansion, one may in principle adiabatically track a many body energy level. Upon adiabatic change of $\Omega$, the levels $\epsilon_i(n)$ for a given subspace $n$ will vary continuously up to, potentially, weak avoided crossings with other levels at the same $n$. However, as long as the expansion is ``convergent,'' there will be no avoided crossings between levels with different $n$. Hence the ``branch choice'' $\Omega n$ to the lab frame energy is uniquely defined by adiabatic continuation and thus equal to $\Omega n$, where $n$ remains the rotating frame occupation number. The only way for this integer contribution to change is via resonances between levels with different $n$ which, as has been noted in other works on high frequency expansions \cite{Bukov2016,Weinberg2017} and described further below, is precisely the mechanism by which the high frequency expansion breaks down.

A detailed proof which confirms this argument is available in Sec.~\ref{sec:proof_branch_choice} for the interested reader.

\subsection{Qudit truncation}
Define the operators
\begin{align}
  \sigma^+ &= \sum^{d-1}_{n=1} |n\rangle\langle n-1| & \sigma^- &= (\sigma^+)^\dagger &	\hat{n} &= \sum^{d-1}_{n=0} n |n\rangle\langle n|,
  \label{eq:operator_qudit}
\end{align} from which one obtains the relations: $\sigma^+\sigma^- = 1 - |0\rangle\langle 0|$, $\sigma^-\sigma^+ = 1 - |d-1\rangle\langle d-1|$, and $\comm{\hat{n}}{\sigma^+} = \sigma^+$.

To obtain a $d/\sqrt{L}$ scaling, let us start by examining the effective rotating frame Hamiltonian:
\begin{align}
  H^{\text{eff}}_{\text{rot}} &= H_0 + \frac{(H_1)^2}{\Omega}\comm{\sigma^+}{\sigma^-} + \frac{\comm{\comm{H_1}{H_0}\sigma^+}{H_1\sigma^-}+\hc}{2\Omega^2} + \ldots \nonumber \\
  &= H_0 + \frac{(H_1)^2}{\Omega}\left( |d-1\rangle\langle d-1| - |0\rangle\langle 0| \right) + \frac{\comm{\comm{H_1}{H_0}}{H_1} \sigma^+ \sigma^- + H_1\comm{H_1}{H_0}\comm{\sigma^+}{\sigma^-} +\hc}{2\Omega^2} + \ldots \nonumber \\
  &= H_0 + \frac{\comm{\comm{H_1}{H_0}}{H_1} (1 - |0\rangle\langle 0|)}{\Omega^2} \nonumber  \\
  &\phantom{=} + \left(\frac{(H_1)^2}{\Omega} + \frac{\comm{H_1}{\comm{H_1}{H_0}}}{2\Omega^2} \right)\left( |d-1\rangle\langle d-1| - |0\rangle\langle 0| \right) + \ldots
  \label{eq:Heffrot_qudit}
\end{align}

Already at low orders we see where the expected all-to-all coupling of the spin chain mediated by the central qudit appears: states with nonzero occupation of qudit states $|0\rangle$ or $|d-1\rangle$ introduce terms proportional to $(H_1)^2 / \Omega$ in $H_\mathrm{rot}^\mathrm{eff}$. Meanwhile, all other terms in Eq.~\ref{eq:Heffrot_qudit} involve only local interactions between spins in the chain due to the nested commutator structure. These local terms consist of the time-averaged Hamiltonian, $H_0$,  which features an MBL-ETH transition, dressed by increasingly non-local terms at higher order in $\Omega^{-1}$. 

Let us now consider what happens to $H_\mathrm{rot}^\mathrm{eff}$ for $\Gamma < \Gamma_c$, where $\Gamma_c$ is the critical value in the Floquet limit. The condition $\Gamma < \Gamma_c$ means that the effective Hamiltonian for the ``infinite temperature state'' $n \sim (d-1)/2$ is localized, which we have just argued corresponds to MBL in a dressed version of $H_0$. On the other hand, even for these values of $\Gamma$, the effective Hamiltonian for $n=0$ has the potential to delocalize, as it involves the competition of a local MBL Hamiltonian $H_0$ against an infinite-range Hamiltonian $(H_1)^2 / \Omega$, with which it does not commute. The question, then, is under what condition will these delocalized $n=0$ eigenstates ``poison'' the localized eigenstates near $n=(d-1)/2$?

\begin{figure}
  \includegraphics[width=.9\textwidth]{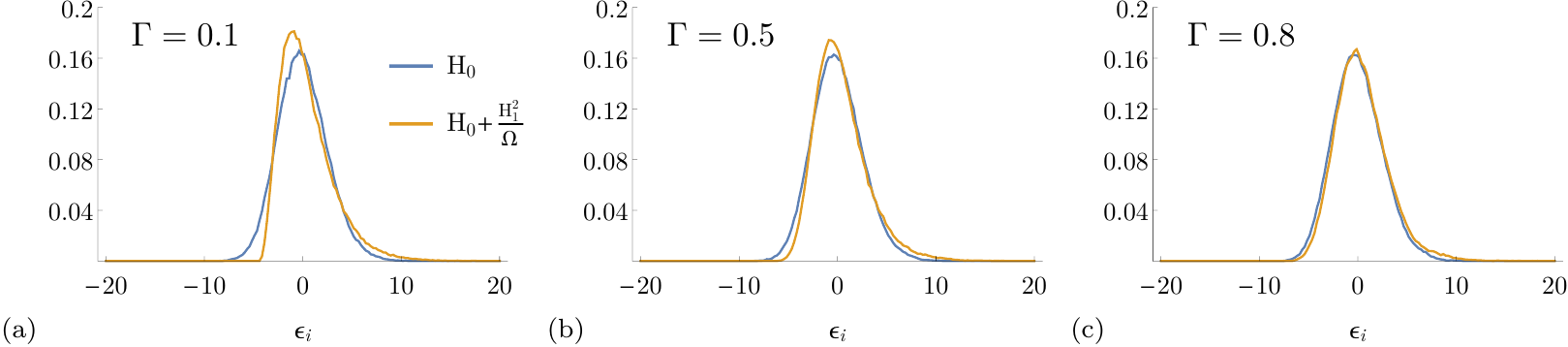}
  \caption{Comparison of many-body density of states (averaged over 100 disorder configurations) for $\Omega=3.927$ and $L=10$ (same parameters as the main body of the paper), showing that the energy scale $J$ controlling the bandwidth mostly remains unchanged even when the Hamiltonian is nonlocal. The distributions remain roughly Gaussian even as the system is tuned with $\Gamma$.}
  \label{fig:DOScomp}
\end{figure}

The density of states for any fixed qudit state $|n\rangle$ is generically given by a Gaussian form \cite{Keating2015, Wilming2017} \[ D_n(E) \sim \frac{2^L}{\sqrt{2\pi} J\sqrt{L}} \exp\left( -\frac{1}{2} \left( \frac{E - \langle E \rangle_n}{J\sqrt{L}} \right)^2 \right), \] where $\langle E \rangle_n \approx n \Omega$ is the average energy for that qudit number. The middle of the spectrum, which we probe, is at $E\approx (d-1)\Omega / 2$. We expect that the states at the edge of the spectrum ($n=0$ and $d-1$) will play no role when their density of states at this energy is much less than that of qudit states near $n=(d-1)/2$, i.e.,
\begin{align*} 
  D_{(d-1)/2}\left(E=(d-1)\Omega/2\right) &\gg D_{0}\left(E=(d-1)\Omega/2\right) \\
  1 &\gg \exp\left( -\frac{1}{2} \left( \frac{(d-1)\Omega/2}{J\sqrt{L}} \right)^2 \right).
\end{align*}
We see that this recovers the $d/\sqrt{L}$ scaling as argued before, and indeed gives a slightly more descriptive scaling $\sim d\Omega/(J\sqrt L)$ in which the energy scales have been restored. Note that, by the high frequency approximation and at moderate system sizes, the density of states $D_0(E)$ for eigenstates with $n=0$ should still be approximately equal to a Gaussian centered around $E=0$ despite the addition of a positive term $(H_1)^2/\Omega$ because this nonlocal term is weak ($J^2/\Omega \ll 1$). This intuition is confirmed via comparing the many body density of states for physically relevant parameters between $H_0$ and $H_0 + (H_1)^2/\Omega$ (Fig.~\ref{fig:DOScomp}).

When $d\lae\sqrt{L}$, the HFE predicts that some delocalized states with $n=0$ and $d-1$ appear at the same energies as the localized states with $n \approx (d-1)/2$. Physically, we predict that these levels would not simply coexist, but rather hybridize via these Floquet resonances \cite{Bukov2016}, which are generally known to break down the HFE. This resonant breakdown in turn destroys the commutation of $H^{\text{eff}}_{\text{rot}}$ and $\hat{n}$, invalidating this crucial assumption used to derive \eqref{eq:framesEnergy}. A full understanding of this non-perturbative regime is beyond the scope of this work (and indeed not fully understood for even much simpler Floquet systems), but nevertheless the crossover is seen to indeed be given by the scaling $d \sim \sqrt L$. 

The above argument is valid for small $L$, for which we indeed see an approximately Gaussian DOS due to the nonlocal term $(H_1)^2/\Omega$ being of similar magnitude to the local term $H_0$. This is no longer true in the thermodynamic limit, as the nonlocal term should dominate and cause the DOS to be exponential with all states at negative energy for $n=0$. This would yield many body ``van Hove singularities'' at $E=0$ and $E=(d-1)\Omega$ in the absence of hybridization between states of different $n$, i.e., when HFE converges. Ultimately, resonances between $n=0$ ($n=d-1$) states with $n=1$ ($n=d-2$) states round out the sharply peaked features, producing thermal states. The likelihood of these thermal states hybridizing with the $n=(d-1)/2$ states will again be small for large $d$, a fact which we believe plays a central role in producing the $d/\sqrt{L}$ scaling. Note that in the absence of hybridization in the thermodynamic limit --- that is, when $(H_1)^2/\Omega$ is dominant --- states with $n=0$ and $d-1$ may themselves be MBL for small enough $\Gamma$ since they are eigenstates of the localizing Hamiltonian $H_1$. The qudit then takes no part in the breakdown of MBL. This picture ceases to be applicable when one needs to worry about resonances, for which a more suitable argument would likely follow along the lines of the ``avalanche'' processes described in \cite{Ponte2017}. Finally, we note that while this analysis involving the nonlocal squared Hamiltonian $H_1$ is more subtle than, say, the photon regularization, the presence of the $H_1^2/\Omega$ term pushes $n=0$ states \emph{downward} in energy and potentially enables MBL in this tail, and thus favors rather than disfavors MBL in the middle of the many-body spectrum. As our numerics seem to rule out MBL stability for a more slowly-increasing qudit size ($d \sim L^\alpha$ for $\alpha < 1/2$) and as the above arguments suggest that $d \sim L^\alpha$ should certainly maintain stability for $\alpha > 1/2$, we again conclude that $d \sim \sqrt L$ is the appropriate scaling form.

A surprising result from this analysis shows that states close to $E=0$, which have $n=0$ in the effective rotating frame, should be delocalized due to competition with the nonlocal term $(H_1)^2/\Omega$. In contrast, highly excited states in the middle of the spectrum do not experience this all-to-all coupling (at least at high drive frequencies) and may be localized if $\Gamma < \Gamma_c$. This further motivates our brief discussion of the possibility of an inverted mobility edge, but further exploration of this intriguing possibility is left for later work.

\subsection{Photon truncation}
We can also consider the photon truncation as mentioned in the main text. It is implicitly understood that this picture only accurately captures the extended zone Hamiltonian in the limit of large average photon number $N_{ph}$ and when fluctuations in photon number between the infinite temperature states are negligible compared to $N_{ph}$. Thus we consider
\begin{align}
  \sigma^+ &= \frac{1}{\sqrt{N_{ph}}}a^\dagger & \sigma^- &= (\sigma^+)^\dagger & \hat{n} &= a^\dagger a
  \label{eq:operator_photon}
\end{align}

Now, at low order, the HFE is give by
\begin{align}
  H^{\text{eff}}_{\text{rot}} &= H_0 + \frac{(H_1)^2}{\Omega}\comm{\sigma^+}{\sigma^-} + \frac{\comm{\comm{H_1}{H_0}\sigma^+}{H_1\sigma^-}+\hc}{2\Omega^2} + \ldots \nonumber \\
  &= H_0 + \frac{(H_1)^2}{\Omega N_{ph}} + \frac{\comm{\comm{H_1}{H_0}}{H_1} a^\dagger a + H_1\comm{H_1}{H_0} +\hc}{2\Omega^2N_{ph}} + \ldots \nonumber \\
  &= H_0 + \frac{\comm{\comm{H_1}{H_0}}{H_1}}{\Omega^2} \frac{\hat n}{N_{ph}} + \nonumber \\ 
  &\phantom{=}+ \frac{(H_1)^2}{\Omega N_{ph}} + \frac{\comm{H_1}{\comm{H_1}{H_0}}}{2\Omega^2N_{ph}} + \ldots \label{eq:H_eff_rot},
\end{align}
As with the qudit truncation, the $(H_{1})^{2}$ term immediately introduces all-to-all couplings at first order in $\Omega^{-1}$. But the strength of this all-to-all coupling is now suppressed as $1/N_{ph}$. Since the photon number $N_{ph}$ plays a role similar to $d$, this explains our result that $N_{ph}\to\infty$ will be MBL for small $\Gamma$, as this is simply a property of the time-averaged Hamiltonian, $H_{0}$. 
Note that, at this order, the actual photon operators drop out - the only thing left is this dependence
on the average photon number, which we put in by hand. However, the
second order term (and higher terms) help us understand why this is
necessary. We see that a term proportional to $\hat{n}=a^{\dagger}a$
appears, multiplied by nested commutators. The details of the nested
commutators are secondary, but the fact that they are commutators
rather than matrix multiplications means that the term proportional
to $\hat{n}$ will be local and thus simply a perturbative dressing
of $H_{0}$. However, in order for the series to meaningfully converge,
the ratio $\langle\hat{n}\rangle/N_{ph}$ must be order 1. At higher
order in the HFE, it is clear that terms proportional to $\hat{n}^{M}$
will come with a denominator of $N_{ph}^{M^{\prime}}$ for some $M^{\prime}\geq M$,
since each $a$ or $a^{\dagger}$ comes with a $(N_{ph})^{-1/2}$.
Thus we see why this regularization is necessary for getting a meaningful
HFE and note that any HFE will break down for photon states with $\langle\hat{n}\rangle\gg N_{ph}$.
It is possible this unbounded part of the spectrum could render our
results asymptotic and lead to breakdown at some very high order,
similar to what happens due to resonance in generic HFEs (cf. Weinberg
et al. \cite{Weinberg2017}).

The next important question is how the direct long-range coupling
due to the second term in Eq.~\ref{eq:H_eff_rot} will cause the
system to delocalize. Let us assume that we set the occupation of
the photon mode to $\langle\hat{n}\rangle\approx N_{ph}$ in scaling
both $N_{ph}$ and $L$ to infinity, such that the $\sqrt{N_{ph}}$
denominator is not simple a trivial rescaling of the coupling to zero.
After that assumption, the photon number no longer enters, and we
are instead left to consider localization within the long-range coupled
model
\[
H_{\text{rot}}^{\text{eff}}\approx H_{0}+\frac{(H_{1})^{2}}{N_{ph}\Omega}=\sum_{j}\left(H_{0,j}+H_{1,j}\frac{\sum_{j^{\prime}}H_{1,j^{\prime}}}{N_{ph}\Omega}\right).
\]
In this expression, we have suggestively split $H_{0/1}$ into a sum
of local terms $H_{0/1,j}$. Because the term $H_{1,j}$ is coupled to all other sites, at large $L$ we may approximate the second term by an effective
field given self-consistently by the expectation value in the desired
eigenstate. Then 
\[
H_{\text{rot}}^{\text{eff}}\approx\sum_{j}\left(H_{0,j}+H_{1,j}\frac{\langle H_{1}\rangle}{N_{ph}\Omega}\right).
\]
For generic many body states, $\langle H_{1}\rangle$ will be a random
variable with zero mean and standard deviation $\sim J\sqrt{L}$.
Thus we see where the $N_{ph}\sim\sqrt{L}$ scaling comes from. For
$\sqrt{L}\ll N_{ph}$, this second term perturbatively dresses the
MBL Hamiltonian $H_{0}$ by local (self-consistent) fields that are
weak, and thus localization survives. For $\sqrt{L}\gg N_{ph}$, the
second term becomes dominant and this mean field approximation becomes
invalid -- indeed there we might expect to again achieve MBL since
$H_{1}$ can also be MBL and shares eigenstates with $H_{1}^{2}$
(although we should be careful about higher order terms in $1/\Omega$).
When the two are comparable, localization and delocalization compete.
We are unable to derive a complete phase diagram at this time, but
this analytical arguments motivates the phase diagram presented in
our paper, which in turn is consistent with numerical data.

\subsection{Proof of Eqs.~\ref{eq:stateRelation} and \ref{eq:framesEnergy} \label{sec:proof_branch_choice}}

In order to show \eqref{eq:stateRelation}, we first note that the fact that $|\psi_{\text{rot}}(t)\rangle\equiv e^{i\Omega\hat{n}t}|\psi(t)\rangle$
for any wavefunction implies that time evolution in the rotating frame is given by
\[
U(t)=e^{-i\Omega\hat{n}t}U_{\text{rot}}(t)e^{i\Omega\hat{n}t}=e^{-i\Omega\hat{n}t}e^{-iK_{\text{rot}}(t)}e^{-iH_{\text{rot}}^{\text{eff}}t}e^{iK_{\text{rot}}(0)}e^{i\Omega\hat{n}t}.
\]
Consider acting on an energy eigenstate $|E\rangle$ in the lab frame for a time $t=T$. Using the fact that $K_{\text{rot}}(T)=K_{\text{rot}}(0)$,
\begin{align*}
  U(T)|E\rangle & =e^{-iET}|E\rangle=\cancelto{1}{e^{-i\Omega\hat{n}T}}e^{-iK_{rot}(0)}e^{-iH_{\text{rot}}^{\text{eff}}T}\underbrace{e^{iK_{rot}(0)}\cancelto{1}{e^{i\Omega\hat{n}T}}|E\rangle}_{\equiv|\epsilon_i; m\rangle}\\
  & =e^{-iK_{rot}(0)}e^{-iH_{\text{rot}}^{\text{eff}}T}|\epsilon_i; m\rangle.
\end{align*}
Multiplying on the left by $e^{iK_\mathrm{rot}(0)}$ and noting that the same thing will work for arbitrary multiples of $T$, we conclude that
\begin{equation*}
  e^{-iH_\mathrm{rot}^\mathrm{eff} T} |\epsilon_i;m\rangle = e^{-i E t} |\epsilon_i;m\rangle~.
\end{equation*}
Hence $|\epsilon_i; m\rangle$ is an eigenstate of $H_{\text{rot}}^{\text{eff}}$ with energy $\epsilon_i$ equal to $E$ up to an integer multiple of $\Omega$. 

Define $|\varphi_{\epsilon}\rangle \equiv |\epsilon_i; m\rangle$ such that $m|\epsilon;m\rangle = \hat{n}|\epsilon; m\rangle$. Calculating time evolution of a lab frame eigenstate for arbitrary $t$ and using Floquet's theorem, we see that
\begin{align*}
  e^{-i H t} |E\rangle & = e^{-i E t} |E\rangle
  \\ & = e^{-i \Omega \hat{n} t} e^{-i K_\mathrm{rot}(t)} e^{-i H_\mathrm{rot}^\mathrm{eff} t} |\varphi_\epsilon \rangle 
  \\ & = e^{-i \Omega \hat{n} t} e^{-i K_\mathrm{rot}(t)} e^{i \Omega \hat{n} t} e^{-i \Omega \hat{n} t} e^{-i \epsilon_i t} |\varphi_\epsilon \rangle 
  \\ & = e^{-i (\epsilon_i + m \Omega) t} e^{-i \Omega \hat{n} t} e^{-i K_\mathrm{rot}(t)} e^{i \Omega \hat{n} t} |\varphi_\epsilon \rangle~.
\end{align*}
Hence,
\begin{equation}
  e^{-i\left(E-(\epsilon_i+m\Omega)\right)t}|E\rangle  =\underbrace{\left(e^{-i\Omega\hat{n}t}e^{-iK_{rot}(t)}e^{i\Omega\hat{n}t}\right)}_{\equiv F(t)^\dagger}e^{iK_{rot}(0)}|E\rangle\label{eq:KickConj}.
\end{equation}

We proceed by proving that the unitary operator $F(t)$ must be time independent. The relationship between the lab frame Hamiltonian $H$ and the effective Hamiltonian $H_{\text{rot}}^{\text{eff}}$and kick operator $K_{\text{rot}}(t)$ in the rotating frame is given by
\begin{align}
\nonumber
  \overbrace{e^{i\Omega \hat{n}t} H e^{-i\Omega \hat{n} t} - \Omega \hat{n}}^{H_{\text{rot}}(t)} & =e^{-iK_{\text{rot}}(t)}H_{\text{rot}}^{\text{eff}}e^{iK_{\text{rot}}(t)}-ie^{-iK_{\text{rot}}(t)}\frac{\partial}{\partial t}e^{iK_{\text{rot}}(t)}\\
  H - \Omega \hat{n} & =F(t)^{\dagger} H_{\text{rot}}^{\text{eff}} F(t)-iF(t)^{\dagger}\left(e^{-i\Omega\hat{n}t}\frac{\partial e^{iK_{\text{rot}}(t)}}{\partial t}e^{i\Omega\hat{n}t}\right), \label{eq:H_minus_Omega_n}
\end{align}
where we have used that $H_{\text{rot}}^{\text{eff}}$ commutes with $\hat{n}$. The time independence of the left hand side imposes restrictions on the right hand side. One possible solution is to have $F(t)=F(0)=e^{iK_{\text{rot}}(0)}\equiv F_{0}$, for which
\begin{align*}
  \frac{\partial}{\partial t}F(t) & =-i\Omega\hat{n}F(t)+i\Omega F(t)\hat{n}+\left(e^{-i\Omega\hat{n}t}\frac{\partial e^{iK_{\text{rot}}(t)}}{\partial t}e^{i\Omega\hat{n}t}\right) = 0\\
\implies  e^{-i\Omega\hat{n}t}\frac{\partial e^{iK_{\text{rot}}(t)}}{\partial t}e^{i\Omega\hat{n}t} & =\left[i\Omega\hat{n},F(t)\right]=\left[i\Omega\hat{n},e^{iK_{\text{rot}}(0)}\right],
\end{align*}
so the second term in Eq.~\ref{eq:H_minus_Omega_n} loses time dependence. Thus a constant $F(t)=F_0$ removes all the time dependence.

To see that this is the only solution, assume that there exists a time-varying solution $W(t)$. The relation between $H$ and $H_{\text{rot}}^{\text{eff}}$ can be rewritten as
\begin{align*}
  H - \Omega \hat{n} & =W(t)^{\dagger}H_{\text{rot}}^{\text{eff}}W(t)-iW(t)^{\dagger}\left(\frac{\partial}{\partial t}W(t)+\left[i\Omega\hat{n},W(t)\right]\right)\\
  %	& =W(t)^{\dagger}H_{rot}^{eff}W(t)-iW(t)^{\dagger}\dot{W}(t)+i\Omega W(t)^{\dagger}\left[\hat{n},W(t)\right]\\
  %	& =W(t)^{\dagger}H_{rot}^{eff}W(t)-iW(t)^{\dagger}\dot{W}(t)+\Omega W(t)^{\dagger}\hat{n}W(t)-\Omega\hat{n}\\
  H & =W(t)^{\dagger}\left(H_{\text{rot}}^{\text{eff}}+\Omega\hat{n}\right)W(t)-iW(t)^{\dagger}\dot{W}(t),
\end{align*}
We have, in addition, the constant solution $F_{0}$
\begin{align*}
  H & =F_{0}^{\dagger}\left(H_{\text{rot}}^{\text{eff}}+\Omega\hat{n}\right)F_{0}.
\end{align*}
Comparing the two equations, we must have
\begin{align}
  F_{0}^{\dagger}\left(H_{\text{rot}}^{\text{eff}}+\Omega\hat{n}\right)F_{0} & =W(t)^{\dagger}\left(H_{\text{rot}}^{\text{eff}}+\Omega\hat{n}\right)W(t)-iW(t)^{\dagger}\dot{W}(t). \label{eq:WVrel}
\end{align}
At time $t=nT$, $W(nT)=W(0)=F_{0}$ and Eq.~\ref{eq:WVrel} becomes
\begin{align*}
  F_{0}^{\dagger}\left(H_{\text{rot}}^{\text{eff}}+\Omega\hat{n}\right)F_{0} & =F_{0}^{\dagger}\left(H_{\text{rot}}^{\text{eff}}+\Omega\hat{n}\right)F_{0}-iW(nT)^{\dagger}\dot{W}(nT)\\
  0 & =W(nT)^{\dagger}\dot{W}(nT),
\end{align*}
which implies $\dot{W}(nT)=0$. Taking a time derivative of Eq.~\ref{eq:WVrel} and evaluating the expression again at $t=nT$, one sees that $0=W(nT)^{\dagger}\ddot{W}(nT)+\dot{W}(nT)^{\dagger}\dot{W}(nT)=W(nT)^{\dagger}\ddot{W}(nT)\implies0=\ddot{W}(nT)$. Repeating this procedure shows all derivatives of $W(t)$ vanishing at $t=nT$, contradicting the assumption of nonconstant $W(t)$.

Thus we see that the RHS of Eq.~\ref{eq:KickConj} is identically $|E\rangle$. As this is true for all times $t$, the exponent of the LHS of Eq.~\ref{eq:KickConj} must vanish. Thus establishes \eqref{eq:framesEnergy}.

\section{Finite size approximants for MBL-ETH transition}
\begin{figure}[h]
	\includegraphics[width=0.9\textwidth]{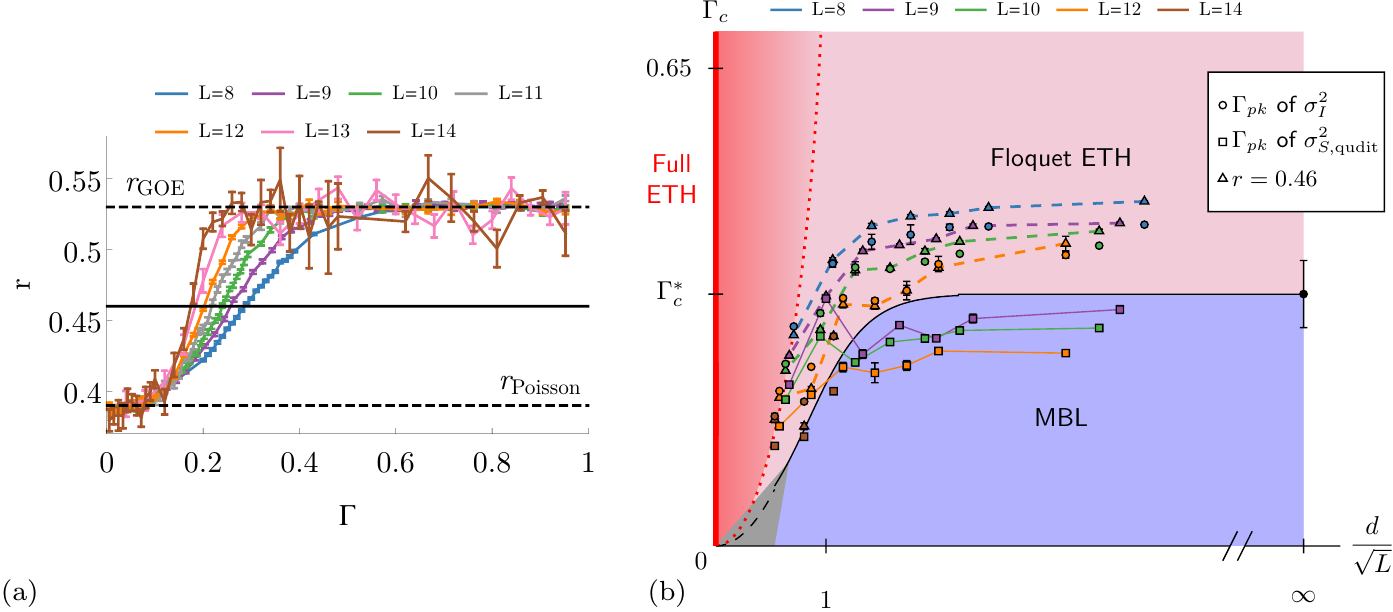}
	\caption{(a) Level statistics ratio $r$ for $d=2$. The curves interpolate between the Poissonian value $r=0.39$ and the Gaussian orthogonal ensemble (GOE) value $r=0.53$. The curves for different $L$ do not appear to cross each other, as one would expect for a finite $\Gamma$ transition. (b) Proposed phase diagram, including finite size estimators for $\Gamma_c$ extracted from the variance of qudit entanglement entropy $S_{\text{qudit}}$ (see main text for details). For ease of identification, data for $\Gamma_{pk}(\sigma^2_{I})$ are connected by dashed lines, whereas data for $\Gamma_{pk}(\sigma^2_{S,\text{qudit}})$ are connected by solid lines.}
	\label{fig:additionalNumerics}
	\end{figure}

An often used metric for determining the integrability of a many-body system is the level statistics ratio $r=\mathrm{min}(\Delta_{n+1},\Delta_n)/\mathrm{max}(\Delta_{n+1},\Delta_n)$, where $\Delta_n=E_n-E_{n-1}$ with sorted energy levels $E_n > E_{n-1}$ \cite{Oganesyan2007}. In MBL systems, this quantity is postulated to have a sharp transition between its extremal values: $r\approx 0.53 (0.39)$ for energy levels distributed according to the Gaussian orthogonal ensemble (Poisson distribution) as the disorder strength $\Gamma$ is tuned. Hence, a good estimate for the critical $\Gamma$ is the location at which the $r$ curves take on a value intermediate between $r = 0.39$ and $0.53$. We find that using $r = 0.46$ as the criterion (see Figure \ref{fig:additionalNumerics}a) gives estimates of $\Gamma_c$ closely matching those made using fluctuations of mutual information variance, $\Gamma_{pk}(\sigma^2_I)$ (see main text). 

Additionally, we can extract the $\Gamma_{pk}$ from the shot-to-shot fluctuations of the qudit's entanglement entropy with the spins, $S_{\text{qudit}}$ (insets of Figure 3 
%\ref{fig:qdsavg} 
of the main text). These values are plotted in Figure \ref{fig:additionalNumerics}b. Particularly, when the spins and the qudit do not thermalize together, $\Gamma_{pk}(\sigma^2_{I})$ is not expected to coincide with $\Gamma_{pk}(\sigma^2_{S,\text{qudit}})$. This corresponds to large $d/\sqrt{L}$ on the proposed phase diagram. In this region, we cannot rule out the possibility for $\Gamma_{pk}(\sigma^2_{S,\text{qudit}})$ converging to $0$ in the thermodynamic limit. For $d/\sqrt{L} \lae 1$ all three estimators appear to converge. This leads us to postulate concurrent thermalization of both the spins and the qudit.

\section{Scaling of mutual information}
We see that the behavior of the half-chain mutual information (MI) mirrors that of entanglement entropy in other studies of MBL: it obeys an area law in the MBL phase and a volume law in the thermal phase. The scaling of MI in the thermal phase is given by the Page value, $\frac{L}{2} - \frac{1}{2\log 2}$. This informs us on the proper scaling form for the variance of MI, which cannot grow at a faster rate -- as a function of system size -- than the Page value. In fact, we see that the peak value of $\sigma_{I}$ grows sublinearly for small $d$ at the system sizes studied. 

\begin{figure}[h]
  \centering
  \includegraphics[width=0.9\textwidth]{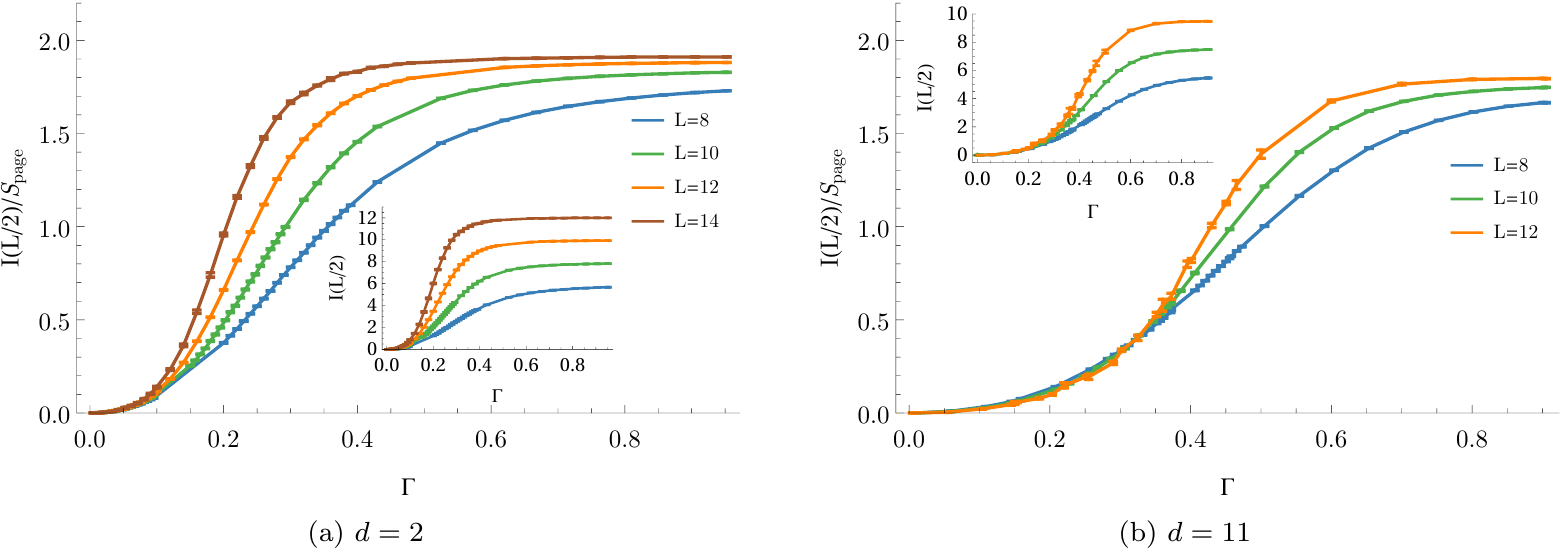}
  \caption{Half-chain mutual information for $d=11$. The unscaled mutual information $I$ (inset) displays area law behavior in the MBL phase (small $\Gamma$). In (b), slow drift in the crossings of scaled $I(L/2)$ are consistent with the slow drift in the peaks of MI variance seen in the inset of Figure 2.
%\ref{fig:MIslowDrift}.
}
  \label{fig:MIavg}
\end{figure}

As stated in the main text, we believe that MBL is absent at $\Gamma>0$ for fixed $d$ in the thermodynamic limit. This is possibly corroborated by the behavior of the normalized MI (Figure \ref{fig:MIavg}) for small qudit sizes where we can probe the $d < \sqrt{L}$ regime. The absence of clear crossings in \ref{fig:MIavg}(a) may indicate a lack of singular behavior in the observable in the thermodynamic limit. This is consistent with the claim that the MBL transition happens at $\Gamma = 0$ for qubit central spins \cite{Ponte2017}. The MI for $d=3$ behaves similarly, leading us to conjecture the full ETH phase at $\Gamma>0$ for $d/\sqrt{L} \to 0$. The limiting value of half-chain mutual information $I(L/2)$ is twice the Page value by definition. Note that this quantity [see equation (2) in the main text] is $I(L/2) = S(A) + S(B) - S(AB)$. In the absence of entanglement between the spin chain and the qudit, $I$ should just equal $2*S(A)$ upon disorder averaging.

\end{document}